\documentclass[namedreferences]{SolarPhysics}
\usepackage[pdfborder={0 0 0 },urlcolor=blue,breaklinks]{hyperref}
\ifx \doiurl \undefined \def \doiurl#1{\href{http://dx.doi.org/#1}{\url{#1}}}\fi
\ifx \adsurl \undefined \def \adsurl#1{\href{http://adsabs.harvard.edu/abs/#1}{\url{#1}}}\fi
\usepackage[natbib,optionalrh,solaenum]{spr-sola-addons} 
\usepackage{graphicx}                    
\usepackage{color}                       
\usepackage{url}                         

\begin{document}

\begin{article}

\begin{opening}
	
\title{Sun-as-a-Star Observation of Flares in Lyman $\alpha$ by the PROBA2/LYRA radiometer}

%
\author{M.~\surname{Kretzschmar$^{1,2}$}, 
	M.~\surname{Dominique$^{1}$},
	I.E.~\surname{Dammasch$^{1}$}
       }

%
\runningauthor{M. Kretzschmar \textit{et al.}}
\runningtitle{Sun-as-a-Star observation of Flares in Lyman $\alpha$ by LYRA}

%
  \institute{$^{1}$ Royal Observatory of Belgium / SIDC, 3 Av. Circulaire, 1180 Brussels, Belgium \\
             $^{2}$ LPC2E, UMR7328 CNRS / Universit\'e d'Orl\'eans, 3a av. de la recherche scientifique, 45071 Orleans, France email: \href{mailto:matthieu.kretzschmar@cnrs-orleans.fr}{matthieu.kretzschmar@cnrs-orleans.fr}
             }

\begin{abstract}
There are very few reports of flare signatures in the solar irradiance at H \textsc{i}  Lyman $\alpha$ at 121.5 nm, \textit{i.e.} the strongest line of the solar spectrum. The LYRA radiometer onboard PROBA2 has observed several flares for which unambiguous signatures have been found in its Lyman-$\alpha$ channel. Here we present a brief overview of these observations followed by a detailed study of one of them, the M2 flare that occurred on 8 February 2010. For this flare, the flux in the LYRA Lyman-$\alpha$ channel increased by 0.6\%, which represents about twice the energy radiated in the GOES soft X-ray channel and is comparable with the energy radiated in the He \textsc{ii} line at 30.4 nm. The Lyman-$\alpha$ emission represents only a minor part of the total radiated energy of this flare, for which a white-light continuum was detected. Additionally, we found that the Lyman-$\alpha$ flare profile follows the gradual phase but peaks before other wavelengths. This M2 flare was very localized and has a very brief impulsive phase, but more statistics are needed to determine if these factors influence the presence of a Lyman-$\alpha$ flare signal strong enough to appear in the solar irradiance.
\end{abstract}

%

\end{opening}

%
 \section{Introduction}\label{s:intro} 
Solar flares radiate energy at all wavelengths, but are best seen in X-ray and extreme ultraviolet light (EUV) where the contrast is high. However, since the background (\textit{e.g.} quiet Sun) flux is several orders of magnitude stronger at longer wavelengths (\textit{e.g.} in the ultraviolet or visible), the dissipation of the flare energy is actually more efficient in these spectral ranges than in EUV and Soft X-Rays (SXR), despite the fact that events show a much smaller contrast there. \\
Lyman $\alpha$ is the strongest line in the solar spectrum and is of particular importance for its impact on the Earth atmosphere. 
Only a few flares have been previously observed in Lyman $\alpha$, especially in full-Sun flux. The primary reason for this is that, because it is technologically very challenging, only a few instruments have observed the Sun at this wavelength. But even from these instruments, very few Lyman-$\alpha$ flares have been reported. It seems thus legitimate to wonder whether the Lyman-$\alpha$ signature is something specific to a certain kind of flare. To investigate this question, we report on the observations of Lyman-$\alpha$ emission as observed by the PROBA2/LYRA radiometer during flares. \\

Several instruments in space have observed the Sun in pass bands that include the Lyman-$\alpha$ line. SOLSTICE \citep{Rottman:1993lr,Woods:1993qy} and SUSIM \citep{Brueckner:1993fk} onboard UARS monitored the ultraviolet irradiance from 119 nm to about 420 nm, from 1991 to 2005. The main objective of these instruments was to access the variability on mid- and long-term intervals (\textit{e.g.} solar cycle) of the irradiance in that spectral range. The data that are easily accessible consist in daily average spectra, where no flare can be detected. Only one flare has been reported over the 14 years of the lifetime of the UARS mission (which recently burned up during its re-entry in Earth's atmosphere): \cite{Brekke:1996ve} reported an observation from SOLSTICE showing a 6\% increase in the full-Sun Lyman-$\alpha$ flux during an X3 flare on 27 February 1992. The line was observed at the end of the impulsive phase and most of the increase was concentrated in the Lyman-$\alpha$ wings.  \\
More recently, \cite{2004GeoRL..3110802W} report a 20\% increase of the irradiance in the core of the Lyman-$\alpha$ line for the extremely large X17 flare that occurred on 28 October 2003. The broad wings of the line were observed to increase by a factor of two.\\
To our knowledge, these are the only two detections of a Lyman-$\alpha$ flare in full-Sun fluxes prior to the LYRA observations presented in this article.\\

Early spectroscopic observations were reported by \cite{Lemaire:1984uq} for a M2.1 flare and showed a maximum intensity of the Lyman-$\alpha$ line of 2.1 10$^{6}$ erg cm$^{-2}$ s$^{-1}$ sr$^{-1}$, about seven times more than the non-flaring active region intensity.  \\
Onboard SOHO, the two spectrometers SUMER \citep{Wilhelm:1995uq} and UVCS \citep{,Kohl:1995fj} have spectral capabilities that cover Lyman $\alpha$. Because of the sensitivity of SUMER to strong solar flux, the instrument was rarely pointing toward flaring regions, especially for observations in the strong Lyman-$\alpha$ line. Using off-limb scattered light in the Lyman continuum near 91 nm, \cite{Lemaire:2004fk} reported an increase of 70\% in the Sun-as-a-star flux at this wavelength. Using a 1D static solar atmosphere model for the quiet Sun and a flare, they estimated the Lyman-$\alpha$ radiance during the flare to be about 350 times the quiet-Sun one. \\
More recently, \cite{Johnson:2011ul} used UVCS observations of the off-limb corona to detect the scattered light of Lyman $\alpha$ during several flares; they estimated the radiative loss in Lyman $\alpha$ to be in the range  3$\times$10$^{25}$ -- 10$^{27}$erg s$^{-1}$.  \\
The NASA TRACE \citep{Handy:1999kx} satellite had a UV imager that includes the Lyman-$\alpha$ line, although it has very significant contribution from wavelengths near 160 nm. \cite{Rubio-da-Costa:2009lr} used a combination of filters to remove this extra-contribution and study the Lyman-$\alpha$ flare signal. They found the Lyman-$\alpha$  flare radiance to be about 80 times the quiet-Sun radiance. Additionally, they estimated the power in Lyman $\alpha$ to be about 10$^{26}$ erg s$^{-1}$ and to represent less than 10\% of the total power deduced from hard X-rays observations for accelerated electrons.\\

Most of these observations have been made during -- or at the end -- of the impulsive phase, where most of the chromospheric and transition region emission is expected. Beyond the fact that none of the above-mentioned instruments was designed to observe flares in Lyman $\alpha$, it is difficult to determine if the low number of flare observations has a physical origin (\textit{i.e.} low flare contrast in Lyman $\alpha$) or if this can be attributed to instrumental and operational causes (\textit{e.g.} poor duty cycle). It remains that the flare emission in the most intense emission line of the solar spectrum is still poorly known. \\

\section{Overview of Flare Observations in Lyman $\alpha$ by LYRA}

 \begin{figure}    
   \centerline{\includegraphics[width=0.69\textwidth,bb=50 270 550 750]{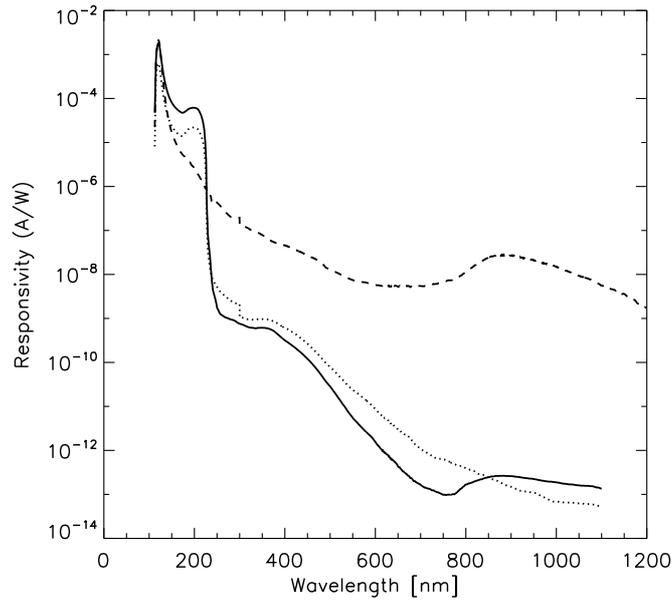}
           }
             \caption{Spectral responsivities (detectors+filters) of the three Lyman-$\alpha$ channels of LYRA. Plain: unit1 ; Dotted: unit2 ; Dashed: unit3.}
   \label{fig_Lyra_Resp}
   \end{figure}

PROBA2/LYRA \citep{Hochedez:2006aa,Dominique:2012} is a radiometer composed of three redundant units. Each of them encompasses the same four broad spectral bands, among which one Lyman-$\alpha$ channel covering the 120 -- 123 nm range. One of these units (unit 2) is used quasi-permanently, while the two other units (units 1 and 3) are used for monitoring the degradation and specific observation campaigns. 

Responsivities (\textit{i.e.} convolution of the filter transmission with the detector response) of all three Lyman-$\alpha$ channels are plotted in Figure \ref{fig_Lyra_Resp}. Unit 3 is provided with a different type of detector than the two other units (silicon \textit{vs} diamond). The responsivity of this unit is therefore slightly different: it does not show any bump around 200 nm, but rejects the longer wavelengths less. In all cases, the rejection of visible and infrared radiation is at most of a few orders of magnitude. This results in a quite strong contamination of the Lyman-$\alpha$ signal by longer wavelengths, to be taken into account when defining the uncertainties on the measurements. 
Of course, this contamination also depends on the emission spectrum of the Sun, which may vary in periods of strong activity.

PROBA2 is orbiting in a helio-synchronous (polar) dawn-dusk orbit, from which the Sun is visible most of the time. This, combined with a very high acquisition cadence (20 Hz nominally), makes the instrument ideal to track flares and analyze their properties. 

Unfortunately, the nominal unit experienced a very strong and rapid degradation, especially in its Lyman-$\alpha$ channel. Only the two first months of operation (January and February 2010) are likely to contain Lyman-$\alpha$ signatures of flares. Therefore, from May 2011, we scheduled several campaigns dedicated to flare observation with unit 3.

After two years in orbit, LYRA has detected signatures of about ten flares, which are summarized in Table 1. We note that the large majority of the flares occurred from the same active region: NOAA 1045. However, it is difficult to conclude that the occurrence of strong Lyman-$\alpha$ flare emission was linked to the particular configuration of this active region since the strong degradation of the Lyman-$\alpha$ channel on the nominal unit is  the principal reason for the absence of other observations later in the mission.

%
 \begin{table}
 \caption{Flares with a signature in Lyman $\alpha$ observed by LYRA}
 \begin{tabular}{ccccc}     
\hline
Date & SXR class & Unit & NOAA region & Quality \\
SOL2010-01-20T10:59  & M1.8 & 2 & 1041& medium (detector not stabilized)\\
SOL2010-02-06T07:04	&C4.0		&2	&1045		&good\\
SOL2010-02-06T18:59	&M2.9		&2	&1045		&medium (pointing manoeuvre) \\
SOL2010-02-07T04:52 	&C9.9		&2	&1045		&medium (pointing manoeuvre)\\
SOL2010-02-07T21:15 	&C4.2		&2	&1045		&good\\
SOL2010-02-08T03:58 	&C2.4		&2	&1045		&medium (pointing manoeuvre)\\
SOL2010-02-08T06:06	&C6.8		&2	&1045		&medium (pointing manoeuvre)\\
SOL2010-02-08T13:47 	&M2.0		&2	&1045		&good\\
SOL2010-02-08T21:23	&M1.0		&2	&1045		&medium (faint, occultation)\\
SOL2011-05-29T21:20	&C8.7		&3	&1227		&medium (pointing manoeuvre)\\
SOL2011-09-08T15:46	&M6.7		&3	&1283		&medium (pointing manoeuvre)\\
 \hline
 \end{tabular}
 \end{table}

\section{The M2 flare on 8 February 2010}

\subsection{Overall description}

The flare to be described here in more detail occurred on 8 February 2010 and was categorized as M2 by the GOES spacecraft. The GOES catalog did not associate the flare with any precise active regions but H $\alpha$ images from the Kanzelh\"{o}he Observatory (the H-$\alpha$ movie can be seen at \url{http://cesar.kso.ac.at/flaremovies/ha1m/2010/20100208_11045_1322_full.mpeg}) show that it occurred in active region AR 11045 that produced many other C and M-class flares. \\

\begin{figure}    
   \centerline{\includegraphics[angle=270,width=0.55\textwidth,bb=150 230 470 550]{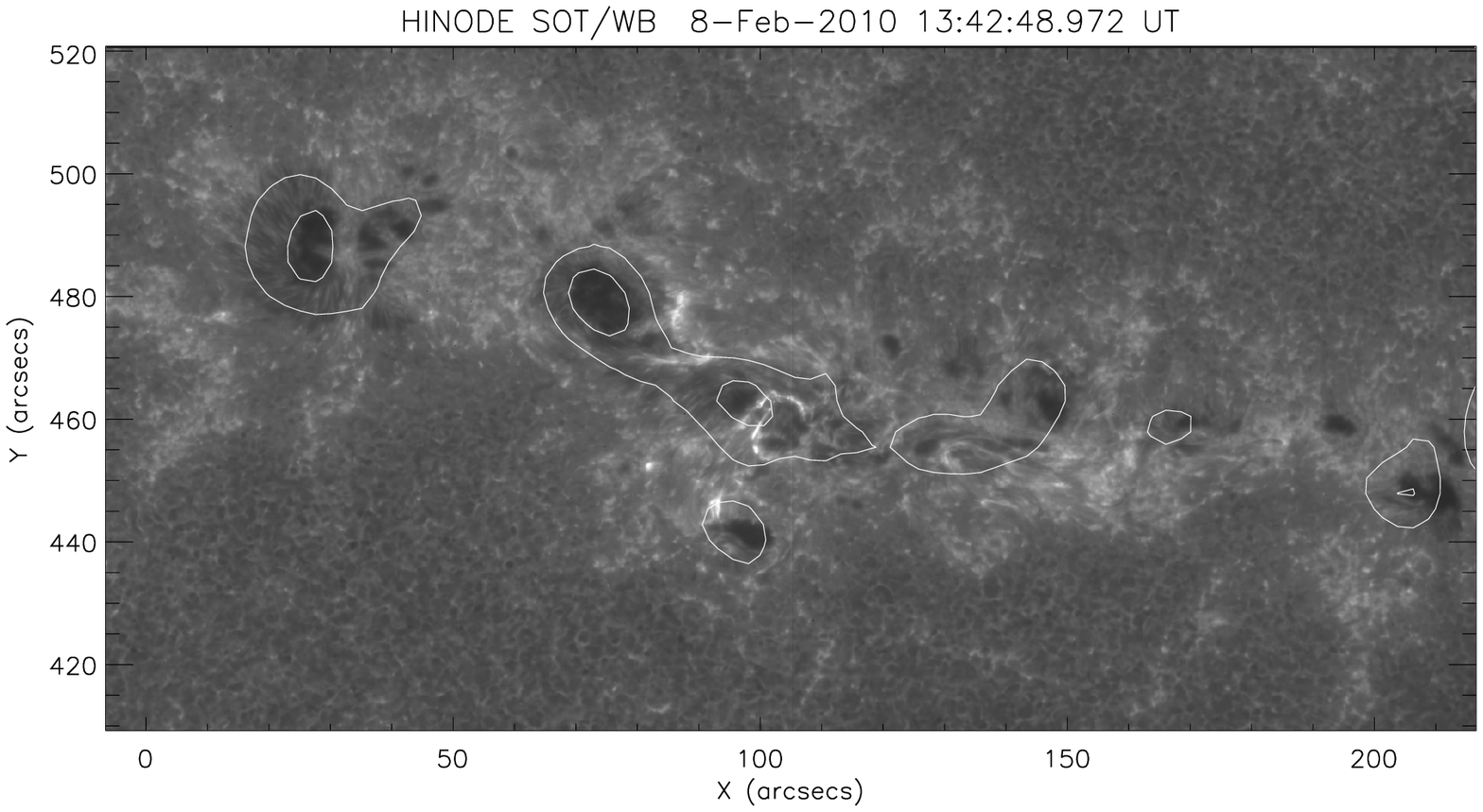}
             }
              \caption{Hinode/SOT images of the active region taken in the Ca \textsc{ii} H line at 396.85nm. Contours correspond to the sunspot position observed on the GONG white-light images.}
   \label{Fig_SOT}
   \end{figure}

Figure \ref{Fig_SOT} shows the position of the sunspot within the active region as observed by GONG, together with a high-resolution image of the chromosphere. The flare happens at $\approx$13:44UT at coordinates ($x$ = 100$''$, $y$ = 46$''$); nothing particular can be seen in the photosphere at this place but the SOT chromospheric images show a rotating structure that indicates the presence of a sheared magnetic field. 
 \begin{figure}    
   \centerline{\includegraphics[angle=90,width=0.55\textwidth,bb=170 170 450 600]{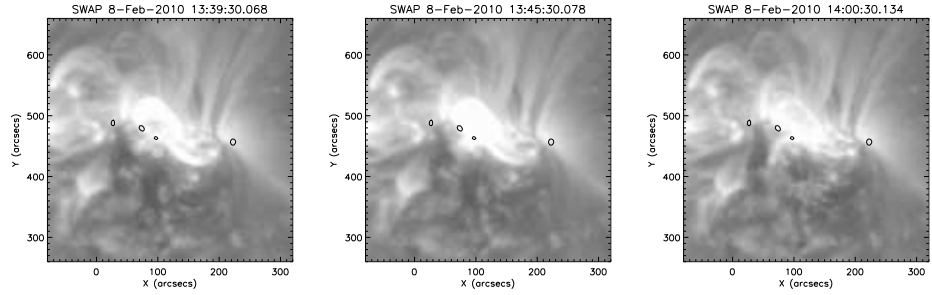}
             }
              \caption{SWAP images of the active region. Contours correspond to the sunspot position observed on the GONG white-light images.}
   \label{Fig_SWAP}
   \end{figure}

To have a look at the higher solar atmosphere, we use the PROBA2/SWAP telescope (\textit{Sun Watcher using Active Pixel System detector and Image Processing}, see \cite{Berghmans:2006lr} and \cite{2012arXiv1208.4631S}). SWAP was observing at a three-minute cadence during the event. Careful examination of the SWAP images (\textit{cf.} Figure \ref{Fig_SWAP}) shows no big modifications of the overall loop system in the active region (AR). The post-flare AR has a thinner loop in its core (or low-altitude loops have disappeared). Only the emission in the very core of the AR is enhanced during the main phase, but saturation of the count rate prevents us from analyzing further the topology in this flaring region. No coronal mass ejection (CME) was detected during the flare. Sub-THz emission has recently been reported for this event \citep{0004-637X-742-2-106}.
 \\
As a whole, the flare results mainly in chromospheric and coronal brightenings without any clear strong modifications of the photospheric magnetic field and coronal loops.\\


\subsection{LYRA Observations}
 \begin{figure}    
   \begin{center}
   \includegraphics[width=0.345\textwidth,angle=90]{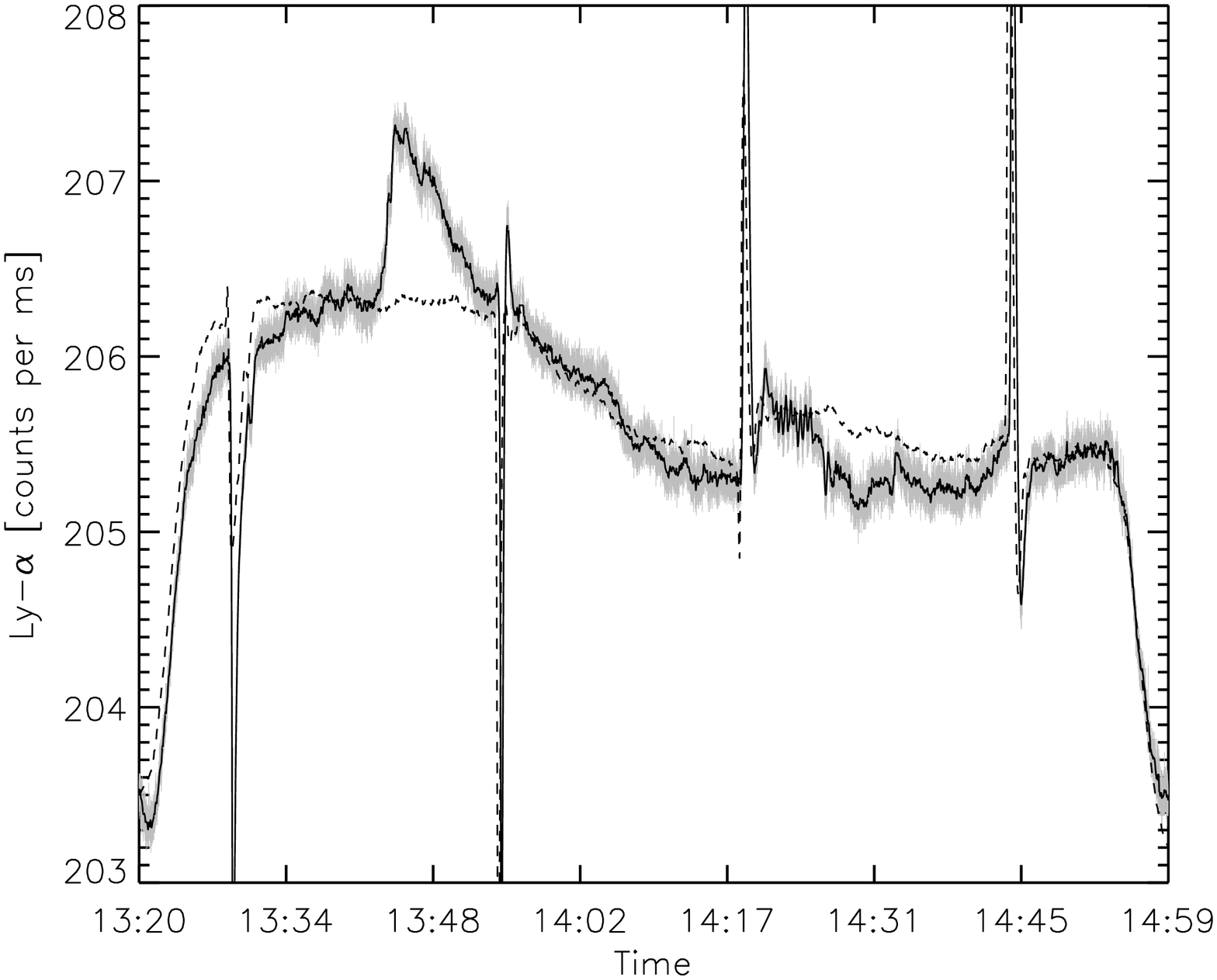} 
   \includegraphics[width=0.345\textwidth,angle=90]{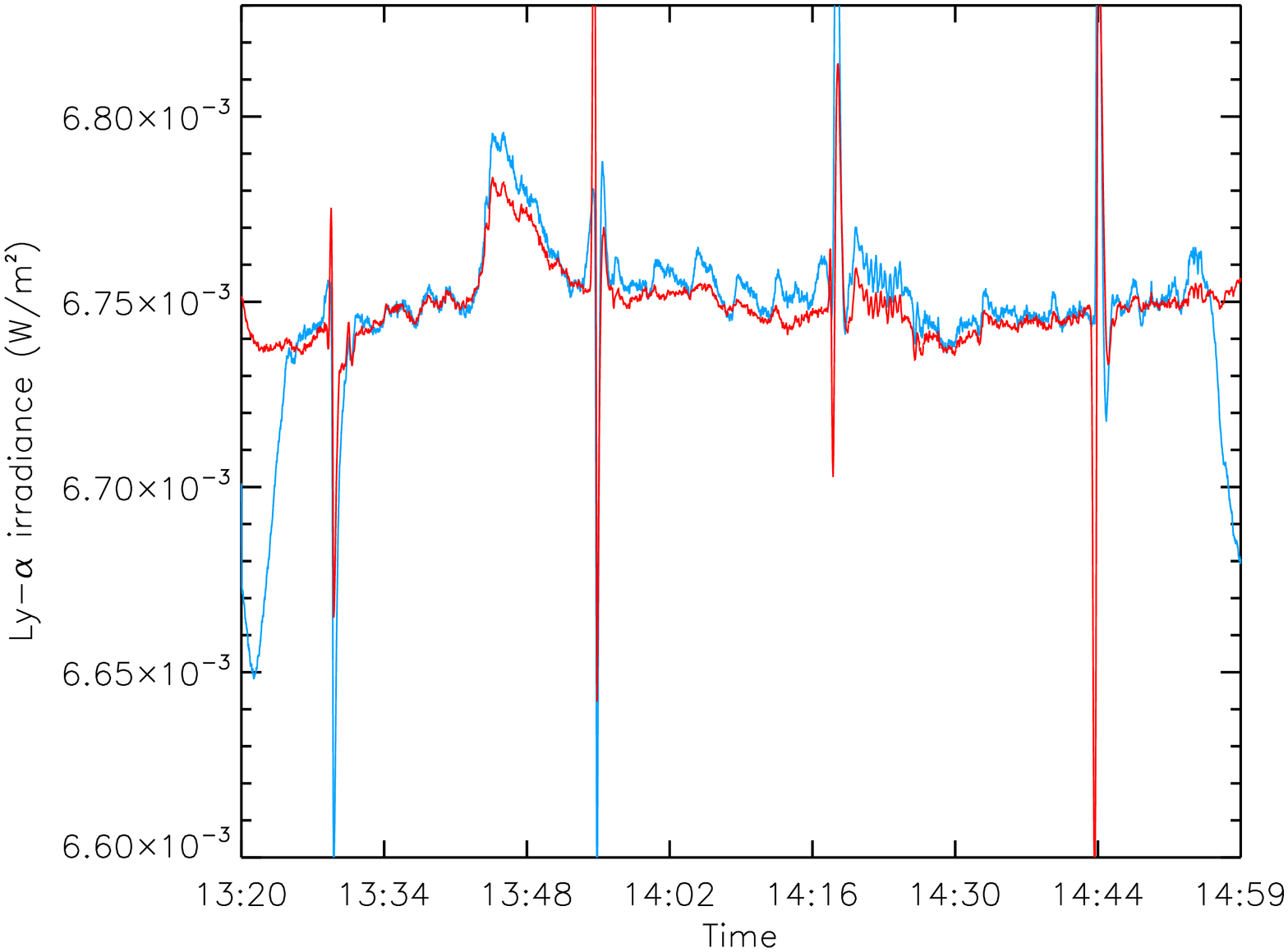} 
              \caption{Left: LYRA channel 1 signal at full (grey) and three-second (black) resolution during the M2 flare of 8 February 2010. The dashed line shows the average flux variation observed over the orbit for that day. Right: LYRA channel 1 signal calibrated following the two ways described in the text: assuming a larger dark current (blue) or using average orbital variations (red).}   \label{Fig_Lya1}
              \end{center}
   \end{figure}
 \begin{figure}    
   \begin{center}
   \includegraphics[width=0.69\textwidth,angle=90]{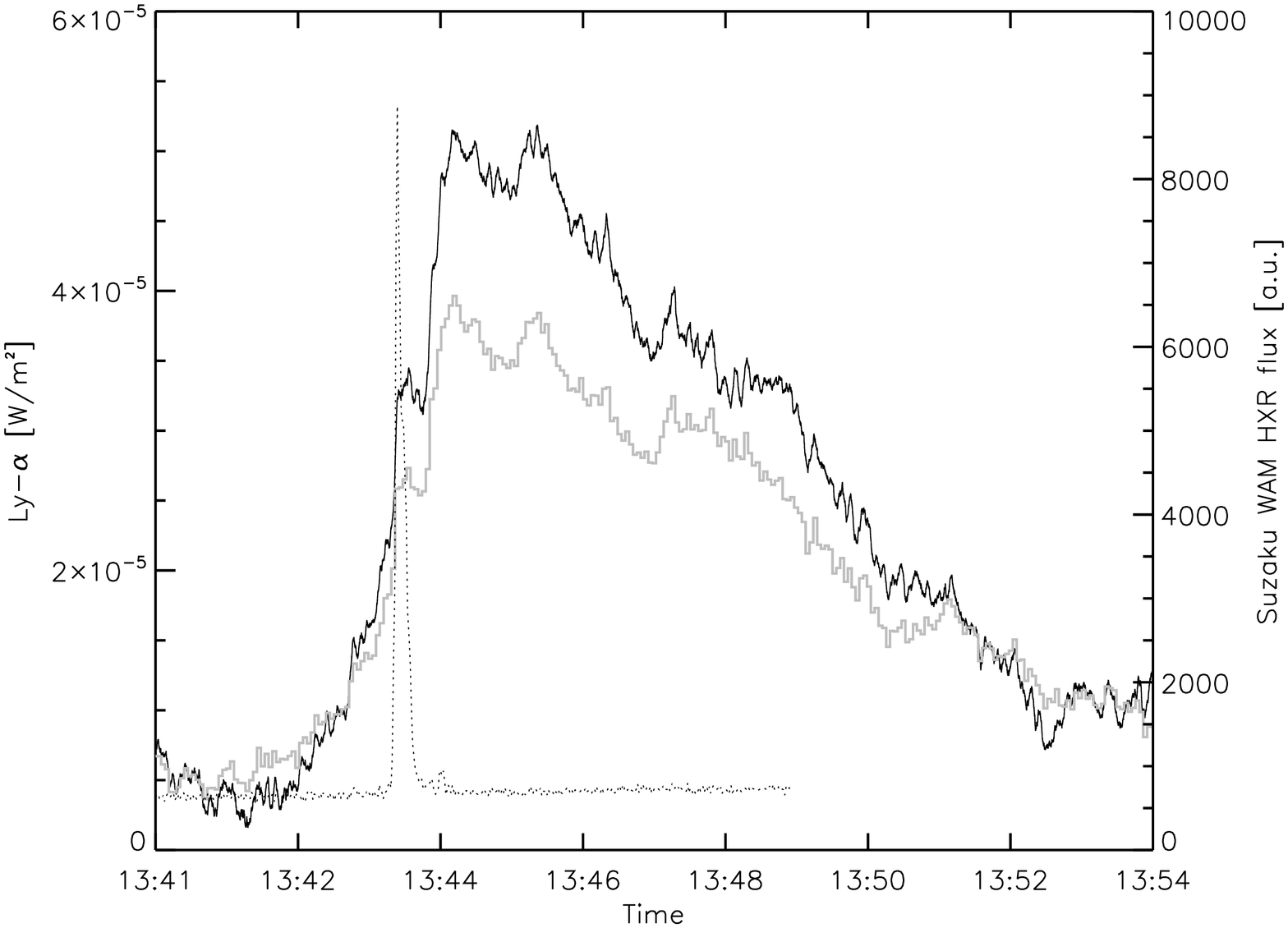} \\
  \hspace{1cm}  \includegraphics[width=0.69\textwidth,angle=90]{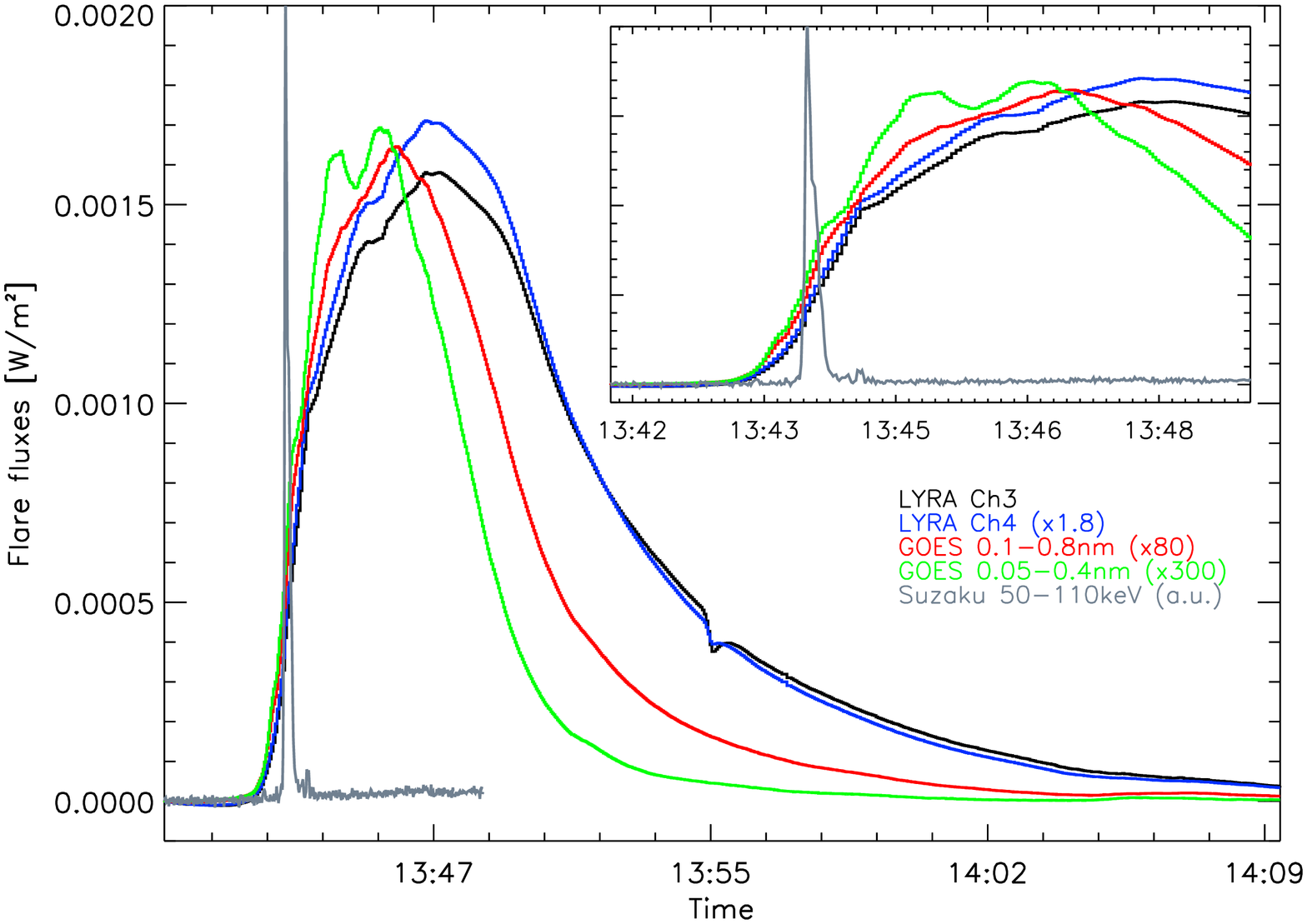}
              \caption{LYRA flare flux during the M2 flare of 8 February 8 2010. Top: Lyman-$\alpha$ channel at 3s resolution, calibrated using the extra dark current correction (black) and the average orbital variations (grey), together with the \textit{Suzaku} hard X-rays flux (dashed line). Bottom: LYRA channel 3 and 4 compared to GOES and Hard X-rays light curves. Pre-flare flux values have been subtracted. The inset shows a zoom on the impulsive phase.}   \label{Fig_Lya2}
              \end{center}
   \end{figure}
%

In this section, we describe in more details the LYRA observations of this flare, as shown in figs.\ref{Fig_Lya1} and \ref{Fig_Lya2}.

\subsubsection{Lyman $\alpha$ Channel}

LYRA level 2 data are calibrated and corrected for degradation. However, because of the way the degradation is corrected (by addition), it is better to use the level 1 data when looking at flares in order to avoid to significantly underestimate the flare increase

This M2 flare occurred in the early days of LYRA, while the instrument was still being tested. Acquisition was still intermittent. In addition, at that period of the year, LYRA was experiencing transits behind the Earth once every orbit. This results in a few minutes of partial extinction of the solar signal, as it is filtered out by the Earth atmosphere. Such an occultation was happening about 25 minutes before the flare (at 13:20, see Figure \ref{Fig_Lya1}), when the LYRA cover was opened after having been closed for 3:20. Nevertheless, when the flare started at 13:40, the spacecraft was back out of the perturbed area, and the signal of the flare should not be affected.\\

The observation was performed with unit 2 (the nominal unit). 
Periodic Large Angle Rotations (LARs) of the PROBA2 spacecraft (approximately every 25 minutes) result in sharp dips and peaks in the time series. This can be seen in Figure \ref{Fig_Lya1} with the four large disturbances occurring at $\approx$13:30, 13:55, 14:19, and 14:45. \\
High-frequency noise affects the original time series (in gray in Figure \ref{Fig_Lya1} left panel) recorded at a cadence of 20 Hz; it has (at least the main part of it) its origin in instrumental effects (see Dominique \textit{et al.}, 2012). For this study, we rebinned the data in three-second intervals to increase the signal-to-noise ratio and match the GOES time series resolution.

Periodic variations of the LYRA channel 1 count rate with the orbit of PROBA2 are observed, as shown in the left panel of Figure \ref{Fig_Lya1}; the overall count-rate trend along the orbit hosting the flare is very well correlated with the average orbit variations. This effect is probably caused by a residual temperature variation (after dark-current subtraction), possibly combined with effects of the inhomogeneous flat field of the detector. The reason why these variations are not observed in the other channels is that channel 1 is amplified by a factor of ten. This orbital variation explains for example why the LYRA signal at the end of the flare (around 14:10) was significantly lower that it was just before the flare. This effect is the major source of uncertainty in the absolute flux value for this event. We deal with this orbital variation in two different ways: First, we assume that it is a pure temperature effect and that for some (unknown) reasons the dark-current correction must be amplified. The right panel of Figure \ref{Fig_Lya1} shows the level-1 data (uncalibrated) after removal of four times the usual dark current estimate; the factor four has been found empirically to reduce at best the orbital variations. The other way, also shown in Figure \ref{Fig_Lya1}, does not make any assumptions on the origin of the orbital variations; they are simply removed by using an average orbital variation computed over the day. In both cases, the conversion in physical units is made by using the SORCE Lyman-$\alpha$ irradiance value for that day (0.00673 W m$^{-2}$) and an additional correction is made to take care of the degradation during that day. \\

As can be seen on Figure \ref{Fig_Lya1}, the first assumption leads to a relative increase of +0.7 \% while the second correction leads to an increase of 0.55 \% . This is significantly less than the 6 \% observed by \cite{Brekke:1996ve}. The difference of a factor of ten can be explained by the fact that the \cite{Brekke:1996ve} flare was an X3 while this flare is an M2 flare. Another explanation could be found in the fact that the LYRA Lyman-$\alpha$ channel has significant contribution from its red wing, which could be less sensitive to flares, thus resulting in an underestimation of the flare flux in the Lyman-$\alpha$ line by LYRA.  \\
It is also interesting to note that the absolute value of the Lyman-$\alpha$ peak flare emission, $\approx$ 4-5 10$^{-5}$ W m$^{-2}$, after background removal (see left panel of Figure \ref{Fig_Lya2}), is about twice the peak value of the\, SXR flux. In the following, we assume that the augmented dark-current correction is the right one, keeping in mind that the actual flare values could be slightly lower.

\subsubsection{EUV and SXR Channels}
Besides the Lyman-$\alpha$ channel, LYRA has three other channels among which two at shorter wavelengths: channel 3 has an aluminum filter and monitors the irradiance between 17 nm -- 80 nm with a contribution below 5 nm; channel 4 has a zirconium filter and covers the range  6 nm -- 20 nm with a contribution below 2 nm. The soft X-ray (SXR) contribution to these channels mainly shows up during flares and these two channels are thus good flare detectors.\\
The right panel of Figure \ref{Fig_Lya2} shows the light curves of LYRA channels 3 and 4 together with the light curves in SXR observed by the two channels of the GOES spacecraft and in hard X-ray (HXR) observed by the japanese spacecraft \textit{Suzaku} (see below). The overall temporal profile of the two LYRA channels is similar to the GOES 0.1 nm -- 0.8 nm, but is slightly delayed in time; this can be explained by the later brightening of the longer ``cooler") EUV wavelengths when the flaring plasma cools down. The fact that lower-temperature emission peaks later has also been well observed by SDO/EVE \citep{Woods:2011rt} and, in this respect, this flare looks ``normal".

\subsection{White Light}
White-light (WL) emission during flares is sometimes observed on solar images. The fact that it is not observed more often is probably due to a lack of temporal and spatial resolution, as well as to poor duty cycles. The WL emission is indeed very intermittent both spatially and temporally \citep[\textit{e.g.}][]{Hudson:2006aa} which makes it observationally challenging. Statistical studies have recently brought significant evidence that basically all flares have WL emission and that it constitutes most (around two thirds) of the energy \citep{Kretzschmar:2010lr,Kretzschmar:2011lr}. The origin of the WL continuum emission during flares is still unclear, although most of the flare energy goes into the continuum. WL flare observations at different pass bands have shown relatively good agreement with a blackbody spectrum at $\approx$9000 K \citep{2007ApJ...656.1187F,Kretzschmar:2011lr}; this is however not supported by basic theoretical considerations, as the flaring atmosphere must be far from thermodynamic equilibrium. Searching for strong WL emission for this event, we did not observe a clear signature of the flare in the total and visible irradiance time series measured by SOHO/VIRGO \citep{1997SoPh..175..267F}, but only a few individual flares show up in these data. However, we found WL emission in the GONG images obtained from the Cerro Tololo Observatory. We now describe their analysis. \\

\begin{figure} 
\centerline{\includegraphics[width=1.3\textwidth,bb=1 5 550 570]{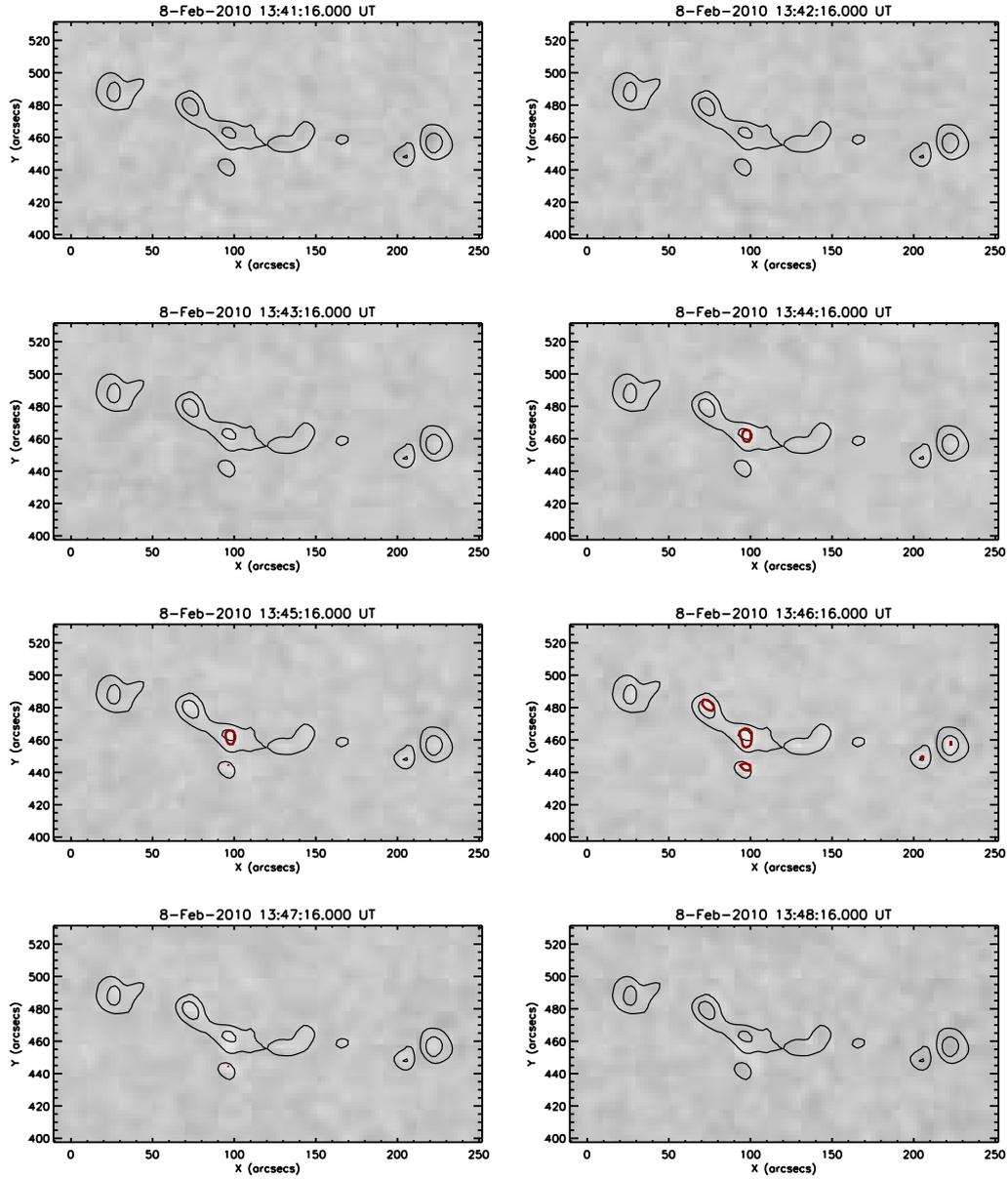}}
 \caption{GONG difference images showing WL emission at the time of the flare (see text for details). The red contour emphasizes regions with at least a 3 \% increase with respect to the median quiet-Sun intensity. Black contours show the position of sunspots.}\label{fig:wlf}
 \end{figure}

The GONG images taken at a one-minute cadence have been carefully co-aligned using cross-correlation. All images have also been normalized to the median quiet-Sun intensity of the image in order to correct for varying seeing conditions and other effects. Finally, we subtracted an average image from each image, so that the images shown in Figure \ref{fig:wlf} correspond to the  excess emission in units of the quiet-Sun (QS) intensity 
$$ \frac{I_{i,j}(t) - <I_{i,j}(t)>_{T}}{I_\mathrm{QS}}$$
assuming that the median quiet-Sun intensity [${I_\mathrm{QS}}$] does not change between images (which is reasonable) and where the period [$T$] used for computing the average image covers about ten minutes before and after the flare.\\

Figure \ref{fig:wlf} shows first WL emission in the core of the active region during the impulsive phase of the flare, starting at the same time as the hard X-rays and high-frequency radio bursts. Later, WL emission appears also elsewhere in the active region, although with a smaller contrast. We attribute this latter WL emission to the gradual phase of the flare. \\

We compute the WL-emission light curves by considering both the emission integrated over all pixels having a contrast larger than 3 \% (which is then named total excess) and the temporal evolution of the one pixel maximal contrast within the whole subfield of view (named maximal excess). These curves are shown in Figure \ref{fig_ts_i} and are commented below. \\

\subsection{Other Wavelengths}
Figure \ref{fig_ts_i} and Figure \ref{fig_ts_imp} show the light curves at several wavelengths during the flare. We describe here how these light curves were obtained (for Lyman $\alpha$ and WL emission we refer to the previous section).\\

\hspace{0.8cm}\emph{Hard X-rays.} RHESSI only observed the late phase of the flare after 13:50, showing thermal radiation at energies below $\approx$12 keV at the end of the gradual phase. The top panels on Figure \ref{fig_ts_i}  and Figure \ref{fig_ts_imp}  show the more energetic HXR emission as observed by the wide-band all-sky camera \citep{Yamaoka:2009lr} onboard the Japanese \textit{Suzaku} spacecraft. This non-thermal emission shows that the impulsive phase of the flare was very brief; simultaneous radio emissions between $\approx$0.8 GHz and  $\approx$4 GHz were also observed by the Ondrejov and Bleien radio observatory (not shown). \\

%
%
 \begin{figure} 
\centerline{
\includegraphics[width=0.5\textwidth,bb=45 50 520 800,angle=90]{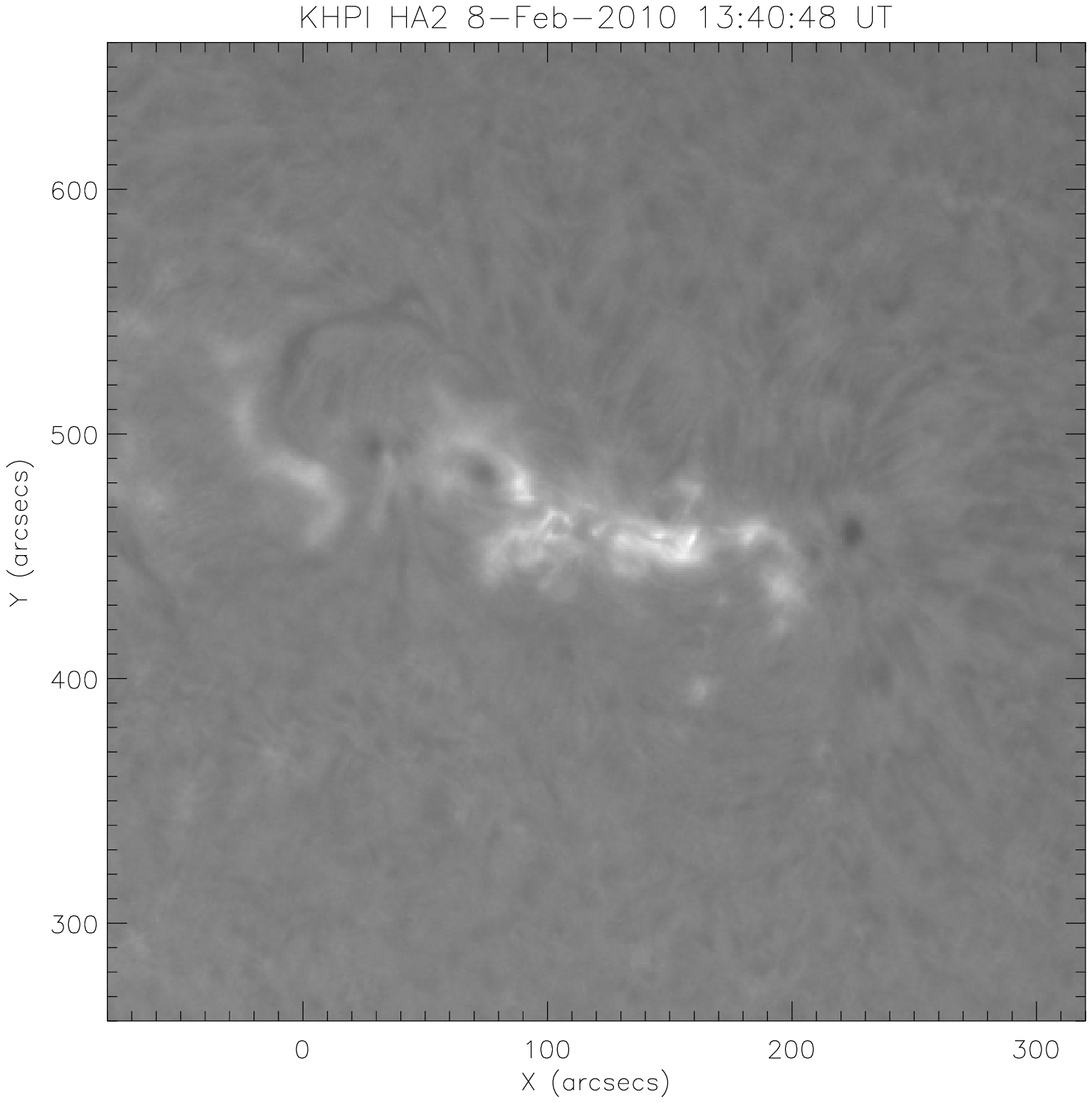}\includegraphics[width=0.5\textwidth,bb=45 50 520 530,angle=90]{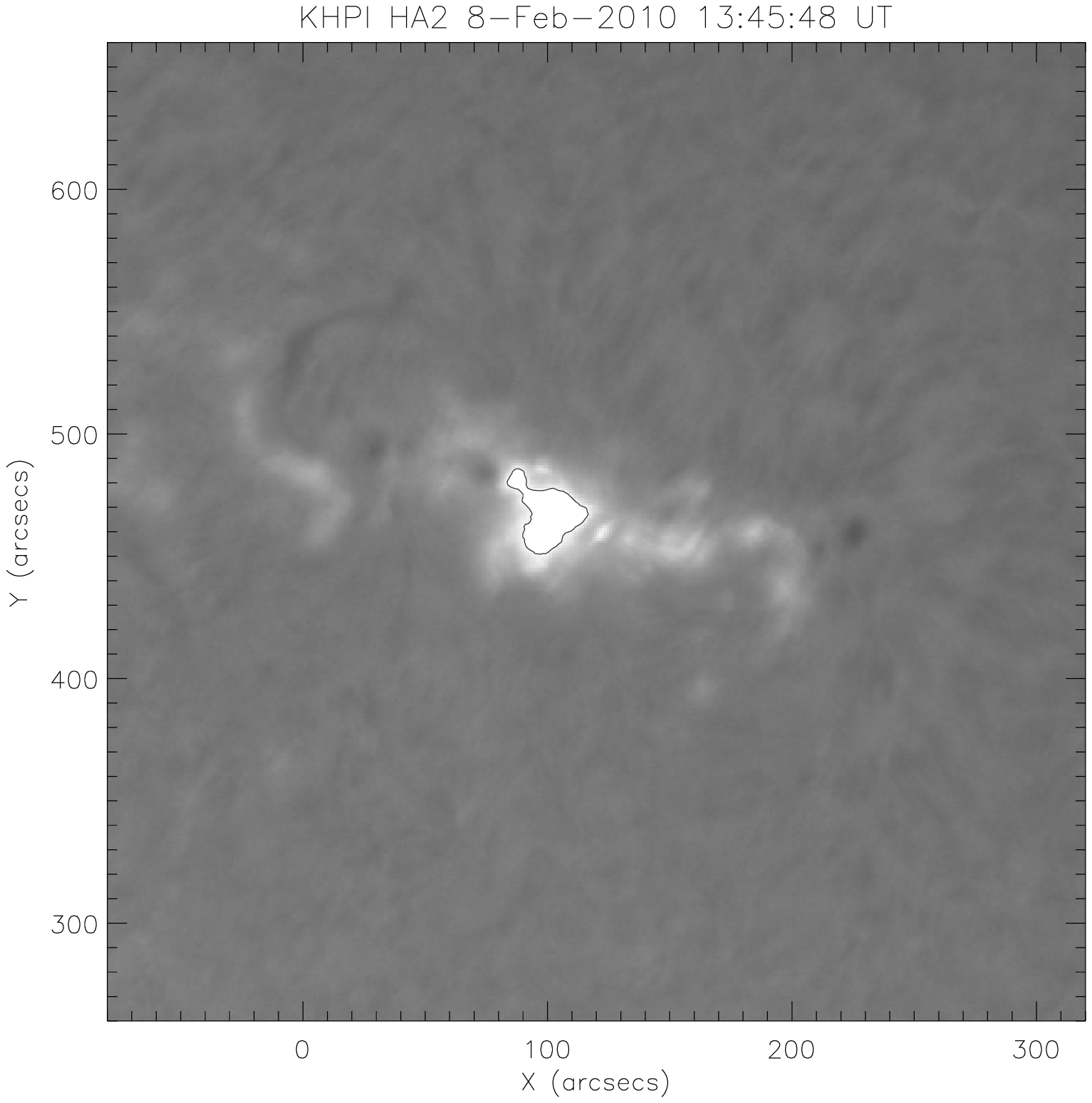} }
\centerline{
\includegraphics[width=0.5\textwidth,bb=45 50 520 800,angle=90]{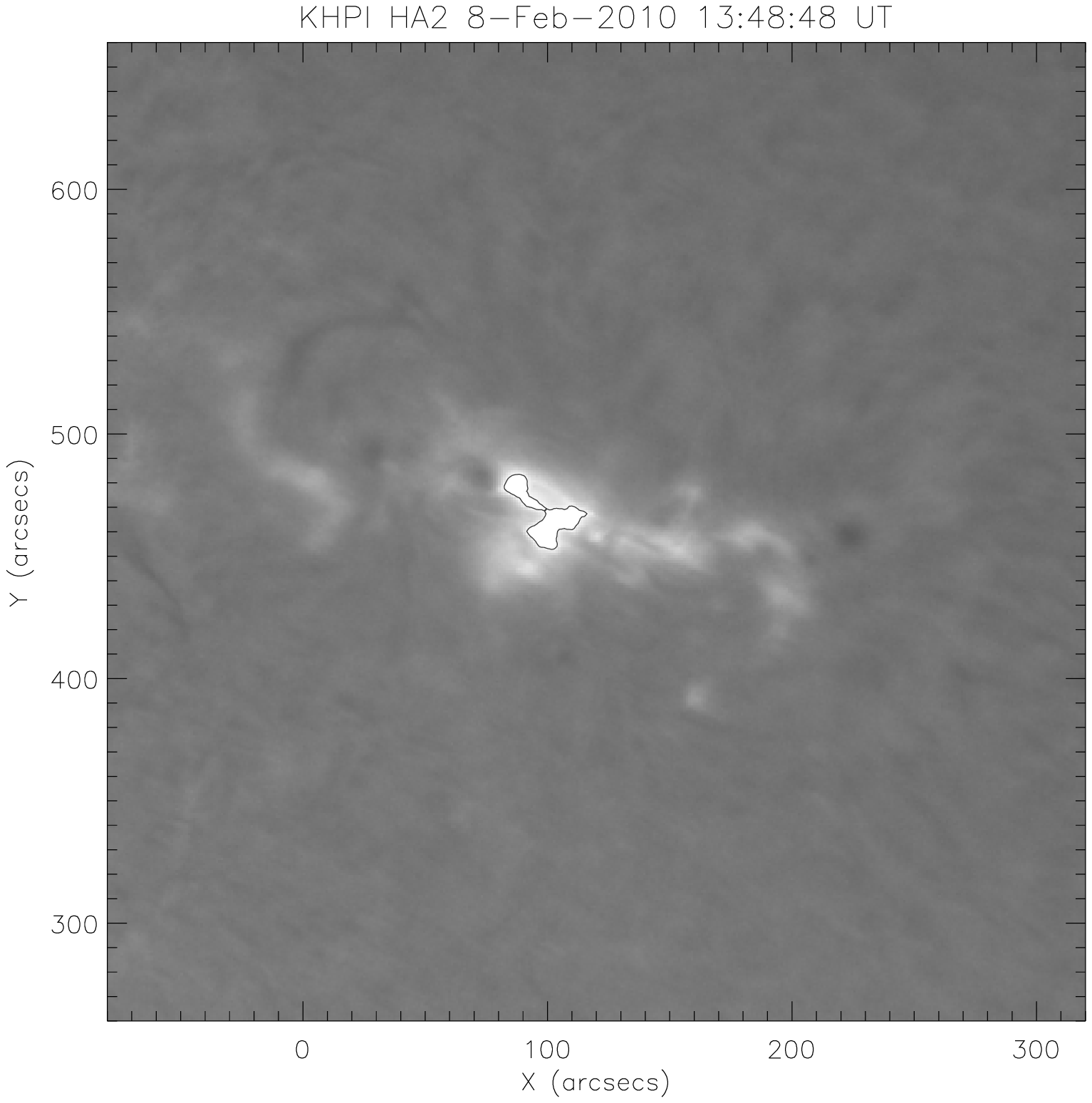}\includegraphics[width=0.5\textwidth,bb=45 50 520 530,angle=90]{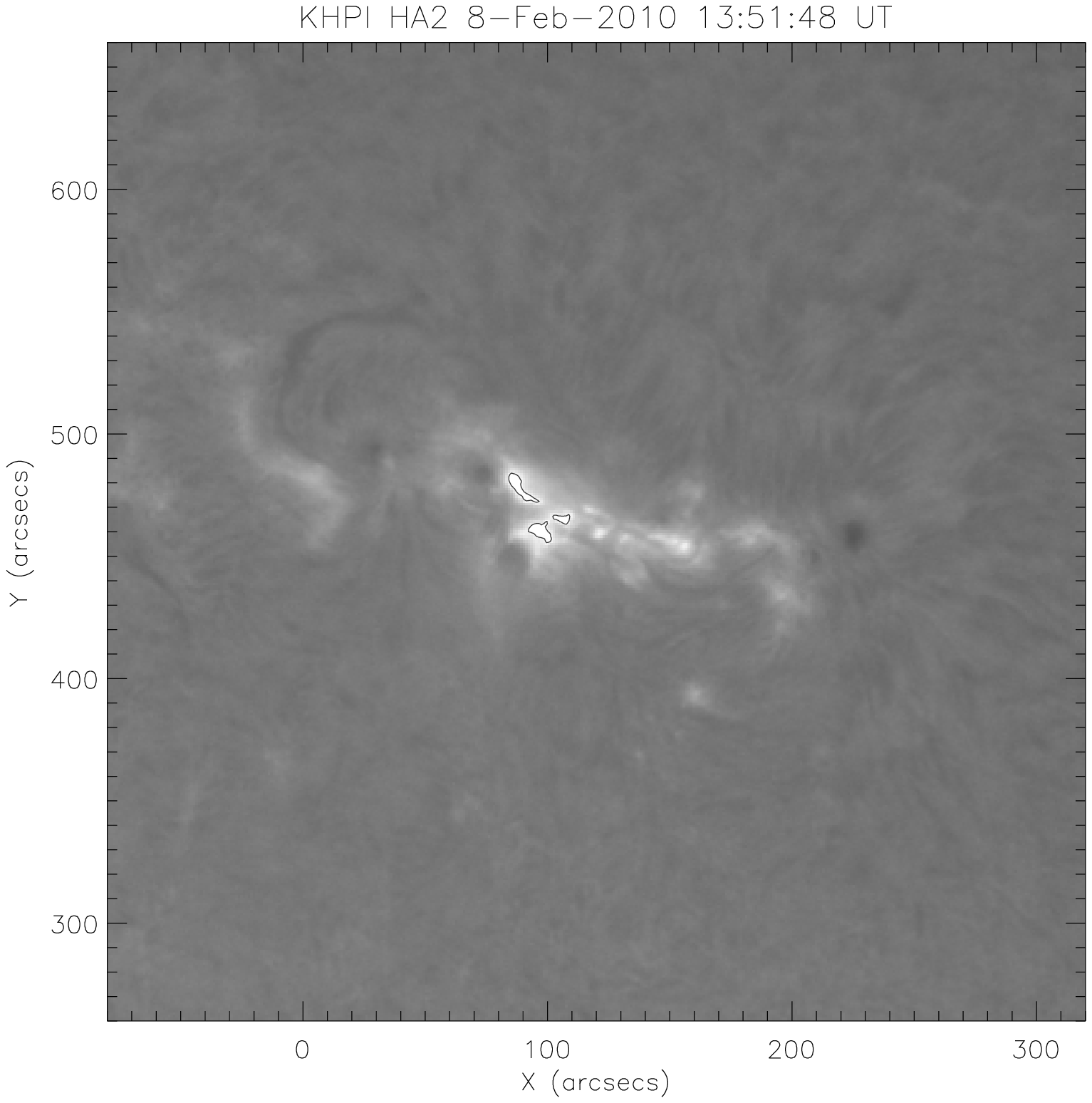}
}
 \caption{H $\alpha$ image normalized to quiet-Sun intensity. The black contour emphasizes regions that saturate.}\label{fig_ha}
 \end{figure}
%
%
\hspace{0.8cm}\emph{H $\alpha$.} The flare has been very well observed in H $\alpha$ by the Kanzelh\"{o}he Observatory (see Figure \ref{fig_ha}). It consists in a localized brightening (details are not accessible due to saturation of counts) within the core of the active region; at the resolution of the observations the H $\alpha$ is co-spatial with the white-light emission. Later on, between 13:55 and 14:05, a filament eruption occurs within the active region. The H $\alpha$ curves shown in Figure \ref{fig_ts_i} and Figure \ref{fig_ts_imp} are deduced from the Kanzelh\"{o}he Observatory; we integrated all pixels whose contrast value is above 1 (which correspond basically to saturating pixel), as illustrated in Figure \ref{fig_ha}.  \\

\hspace{0.8cm}\emph{EUV and SXR flux.} The extreme-ultraviolet light curves come from the SOHO/SEM and GOES instruments. The well-known GOES  observations are in two spectral ranges, from 0.1 nm to 0.8 nm and from 0.05 nm to 0.4 nm. The \textit{Solar Extreme Ultraviolet Monitor} (SEM) is a transmission grating spectro-photometer that measures the solar flux between 0.1 and 50 nm at the central order and between 26nm and 34nm (thus including the strong He \textsc{ii} 30.4 nm line) in first order.\\ 

%
\section{Results and Discussions}
\subsection{Overall Flare Evolution}
%
%
 \begin{figure} 
\centerline{\includegraphics[width=0.8\textwidth,bb=50 40 650 780]{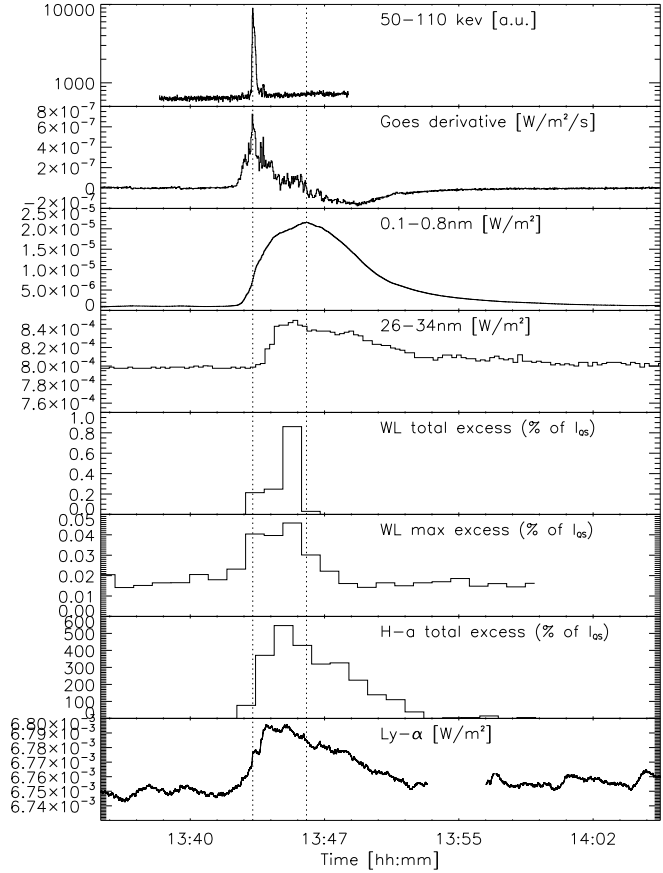}}
 \caption{Light curves at different wavelengths for the M2 flare of 8 February 2010. }\label{fig_ts_i}
 \end{figure}
 \begin{figure} 
\centerline{\includegraphics[width=0.8\textwidth,bb=50 40 650 780,clip=]{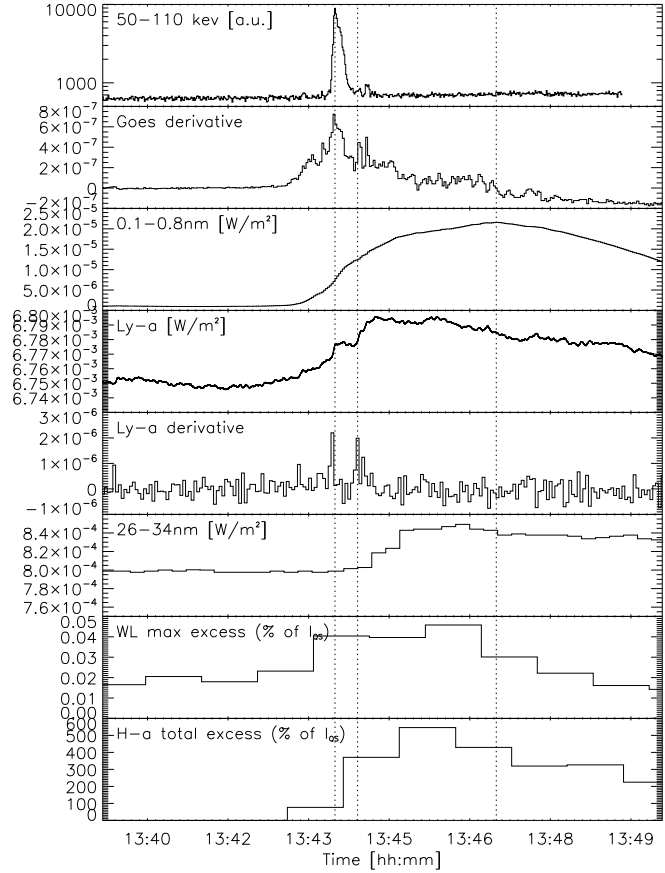}}
 \caption{Zoom on the impulsive phase for the M2 flare of 8 February 2010.}\label{fig_ts_imp}
 \end{figure}
%
%
Several features are interesting to note in inspecting Figures \ref{fig_ts_i} and \ref{fig_ts_imp}. 
The flare has a very brief impulsive phase, of about ten seconds duration.  Figure \ref{fig_ts_imp} shows the occurrence of a second, smaller, hard X-ray burst, about 30 seconds after the first one. This is also well observed in the derivative of the GOES flux as well as, interestingly, in the derivative of the Lyman-$\alpha$ flux. Some emission enhancement can be seen in the GOES and Lyman-$\alpha$ light curves before the HXR burst. This suggests the existence of heating before the appearance of accelerated particles; this could be caused by the dissipation of the first instabilities of the active-region magnetic field just before the magnetic reconnection. This would also be consistent with the scenario proposed by \cite{2008ApJ...675.1645F} where the energy is first transported through waves from the reconnection site to the chromosphere where particles are then accelerated.  \\
WL emission appears during the impulsive phase (with the limiting time sampling of GONG) and lasts for about three minutes. At 13:46, however, the WL emission is well spread over the entire sunspot complex which results in a large increase ($\approx$ a factor four) of the WL total excess; we cannot rule out that part of this emission arising far from the active-region core is actually caused by artifacts, \textit{e.g.} errors in GONG images co-alignement. Indeed, the H-$\alpha$ images do not show comparable emission (see Figure \ref{fig_ha}).\\
The SEM 26 -- 34 nm flux looks delayed by 45 seconds, for which we have no explanation.\\ 
The Lyman-$\alpha$ flux observed by LYRA seems to follow the gradual phase (as illustrated by  the correlation of the Lyman-$\alpha$ derivative and the HXR bursts) but peaks earlier than all other channels. Inspection of the behavior of LYRA channel 3 and 4 on Figure \ref{Fig_Lya2} shows that emission at lower temperature than GOES tend to peak later, as a results of cooling of the hot material. These observations support then the idea that the Lyman-$\alpha$ enhancement takes place together with the initial heating of chromospheric plasma to hotter temperature.

\subsection{Energy Release}
We now try to estimate the energy release by this flare at various wavelengths. Because RHESSI was not observing we cannot estimate the energy in accelerated particles (the wide-band all-sky camera onboard \textit{Suzaku} does not allow us to produce spectra) and we concentrate on the radiative outputs only.
%
\subsubsection{WL Continuum}
In the following, we assume that the flare WL-continuum spectrum is close enough to a Planck distribution, so that we can use the Planck law to retrieve basic energy estimates. This assumption is probably not correct but it has proven to lead to energy estimates that are consistent with the observations. The excess emission due to the flare is faint (see Figure \ref{fig:wlf}), \textit{i.e.} $\approx$3 \% (up to 4.6 \%, and with a mean of 3.6 \%) of the quiet-Sun intensity. A 3.6 \% increase at the GONG wavelength corresponds to a flaring temperature of about 50 K above the photospheric temperature (5777 K) only. If we use this temperature increase and the flaring area determined by the 3.6 \% level, the total energy that goes into WL can be estimated by
$$ n_\mathrm{px}*A_\mathrm{px}*\int_{\lambda}I_\mathrm{planck}(\lambda)\mathrm{d} \lambda $$ 
where $n_\mathrm{px}$ is the number of flaring pixel (above 3.6 \%) and $A_\mathrm{px}$ is the area of one pixel. We find a radiative loss of the order of $\approx 10^{30}$ erg per minute interval ($\approx$1.7 10$^{28}$erg s$^{-1}$). This very rough estimate is in surprisingly good agreement with the results of \cite{Kretzschmar:2011lr}, based on completely different data set  and methods (superposed epoch analysis of irradiance time series, \textit{i.e.} no information on local contrast and flaring area) but which also assumes that the WL-continuum follows a blackbody curve. Let us note, however, that although the WL-energy estimate is consistent for the two approaches, the estimations of the observed flaring surface and contrast are very different.  \\

\subsubsection{Coronal Emission}
 \begin{figure} 
\centerline{\includegraphics[width=0.6\textwidth,bb=120 130 500 600,angle=90]{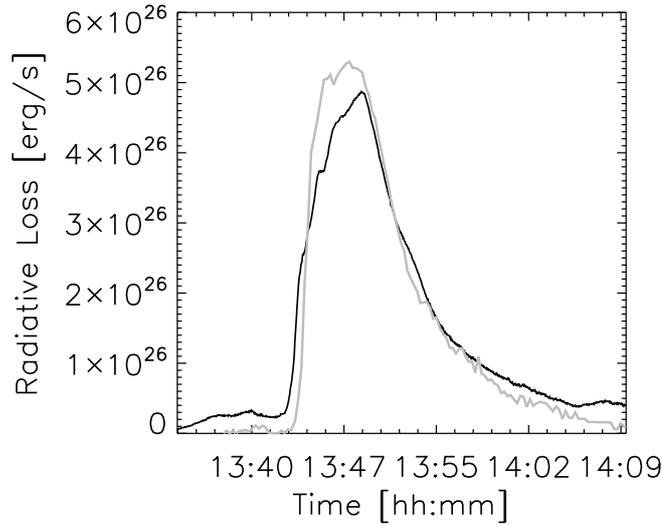}}
 \caption{Radiative loss for the M2 flare of 8 February 2010, as derived from the GOES channels (in black) and the SEM 0.1 -- 50 nm channel (in gray).}\label{fig_CorLoss}
 \end{figure}
Figure \ref{fig_CorLoss} shows the radiative loss as estimated from the two SXR channels of GOES and from the SEM 0.1 -- 50 nm channel. Although the values obtained are very similar, the two estimates have intrinsically several differences: the GOES estimate relies on the computation of the emission measure at one predominant temperature, here computed to be between 10$^{7}$ K and 2$\times 10^{7}$ K during the flare. It thus represents emission at very high temperature and short wavelengths ($\lesssim$ 20 nm) that occurs in the flaring loops. The SEM channel includes also emission at these high temperatures, together with emission of the colder corona and transition region, as well as some chromospheric emission (\textit{e.g.} the He \textsc{ii} line at 30.4 nm); it misses very-short wavelengths corresponding to high temperature emission. The SEM radiative loss is higher during the impulsive phase while the GOES radiative loss has a longer gradual phase, which reflects that there is more chromospheric emission and less hot coronal emission in the SEM channel.  \\
It is difficult to deduce the true coronal loss from these curves, but it seems reasonable to assume that the coronal radiative loss at flare maximum does not exceed 2$\times$10$^{27}$erg s$^{-1}$, \textit{i.e.} five times what is observed in GOES and SEM. This is about eight times less than the estimated power in the WL, without taking into account the fact that EUV and SXR emission stand for a longer time than the WL emission.

\begin{figure} 
\centerline{\includegraphics[width=0.8\textwidth,bb=0 0 600 800,angle=90]{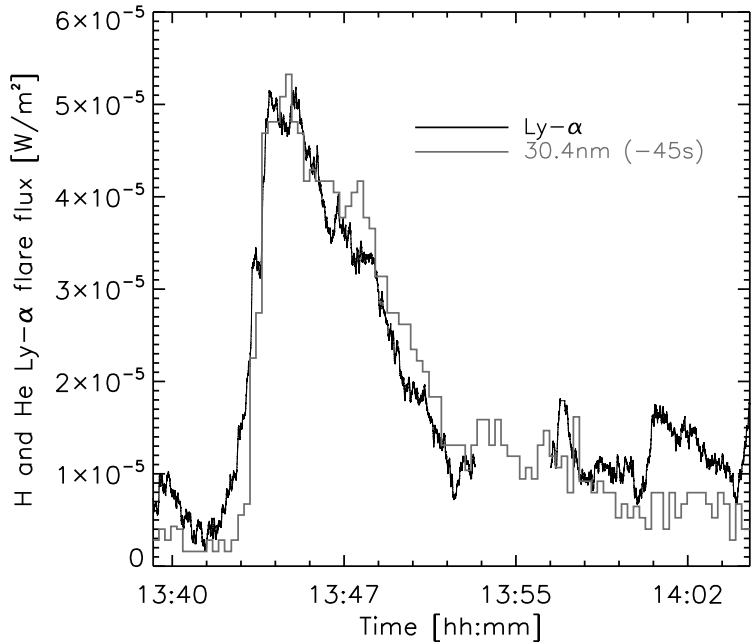}}
 \caption{Light curves for the Lyman-$\alpha$ line of H \textsc{i} and He \textsc{ii} for the M2 flare of 8 February 2010. }\label{fig_lya_he1}
 \end{figure}

\subsubsection{Lyman $\alpha$ and He II 30.4 nm }

Figure \ref{fig_lya_he1} shows the Lyman-$\alpha$ channel flare flux (background subtracted) together with the flare flux measured by SEM around the He \textsc{ii} line at 30.4 nm; for better comparison, this later curve has been translated by -45 seconds. The excellent match of the two profiles suggest a possible error of timing in the SEM data. However, the most relevant point for this work is that the same amount of the flare energy is released in the SEM 30.4 nm channel, which actually covers from 26nm to 34nm and includes the Lyman-$\alpha$ transition of the He II ion at 30.4 nm. \\
The peak increase in the LYRA Lyman-$\alpha$ channel is between  0.5 \% and 0.7 \% and corresponds to an extra 4 -- 5 $\times$10$^{-5}$W m$^{-2}$ with respect to the pre-flare flux. This is about twice the energy release in the GOES 0.1 -- 0.8 nm channel, although the relative increase is much smaller. By assuming an isotropic distribution, we estimate the energy released in Lyman $\alpha$ between 13:42 and 13:54 to be $\approx$5 $\times$10$^{27}$ erg, which corresponds to a mean power of 7$\times$10$^{24}$ erg s$^{-1}$. The mean power at flare peak is 1.4$\times$10$^{25}$ erg s$^{-1}$. These number are significantly less than what has been found previously \citep[\textit{e.g.}][]{Johnson:2011ul,Rubio-da-Costa:2009lr} for comparable events. However, using a flaring area deduced from the H $\alpha$ observations (see Figure \ref{fig_ha}), we found the flare intensity to be within a factor of two to what has been observed by  \cite{1978SoPh...60..341M}, \cite{1980SoPh...67..339C}, and \cite{Lemaire:2004fk} in spectroscopic observations (around 10$^{6}$ erg s$^{-1}$ cm$^{-2}$ sr$^{-1}$). This also agrees with \cite{Rubio-da-Costa:2009lr} if we apply an additional correction of 2.5 to the calibration of the TRACE Lyman-$\alpha$ channel as discussed in their article.\\

\section{Conclusions}

In this work, we have given a quick overview of the flares observed by LYRA in its Lyman-$\alpha$ channel and have performed a detailed study of one event. Most of the flares with Lyman-$\alpha$ emission observed by LYRA come from the same active region. However, the lack of observations of Lyman-$\alpha$ flare signals at later times is more probably due to the strong degradation of the LYRA Lyman-$\alpha$ channel rather than to the specific configuration of this active region. \\
For the M2 flare of 8 February 2010, an increase of 0.5 -- 0.7 \% of the irradiance was measured by the LYRA Lyman-$\alpha$ channel. We have shown that this flare was very localized and did not remarkably affect the active region configuration at the photospheric and coronal levels. This small increase in LYRA Lyman $\alpha$ represents about $10^{25}$ erg s$^{-1}$ and an overall flare emission of $\approx$5$\times$10$^{27}$ erg. This is only a very small portion of the total energy radiated by the flare: coronal radiation has been estimated to represent on the order of $\approx$10$^{26}$erg s$^{-1}$, \textit{i.e.} ten times more; we have also detected the presence of a white-light continuum for this flare and estimated it to be even larger, of the order of 10$^{28}$erg s$ ^{-1}$. \\
The estimated flare emission in Lyman $\alpha$ agrees with the previous studies, within the uncertainties. \\
 A very brief and intense non thermal hard X-ray burst occurred at the beginning of the flare, but some Lyman-$\alpha$ emission seems to start earlier. The Lyman-$\alpha$ curve follows the behavior of other light curves characteristic of the gradual phase (\textit{e.g.} GOES/SXR) but Lyman $\alpha$ peaks earlier than the other curves. This confirms previous results obtained in the rare analyses of the response of the Lyman-$\alpha$ line to flare. \\
Finally, this work demonstrates that LYRA is a suitable instrument to analyze the Lyman-$\alpha$ emission during flares.

\section{Acknowledgment}
LYRA is a project of the Centre Spatial de Li\`ege, the Physikalisch-Meteorologisches Observatorium Davos, and the Royal Observatory of Belgium funded by the Belgian Federal Science Policy Office (BELSPO) and by the Swiss Bundesamt f\"ur Bildung und Wissenschaft. SWAP is a project of the Centre Spatial de Li\`ege and the Royal Observatory of Belgium funded by the Belgian Federal Science Policy Office (BELSPO). This work utilizes data obtained by the Global Oscillation Network Group (GONG) program, managed by the National Solar Observatory, which is operated by AURA, Inc.  under a cooperative agreement with the National Science Foundation. This work has received funding from the European Community's Seventh Framework Programme (FP7/2007-2013) under the grant agreement n¡ 261948 (ATMOP project, www.atmop.eu).

\bibliographystyle{spr-mp-sola} 
\bibliography{Lya_flare_last.bib} 

%

%

%

%

%

%
%

%
%
%
%
%
%

\end{article} 
\end{document}